\documentclass[aps,prl,twocolumn,showpacs,floatfix,superscriptaddress]{revtex4}
\usepackage{graphicx}
\usepackage{dcolumn}
\usepackage{amsmath}
\usepackage{latexsym}

\begin{document}

\title{Effective interactions and melting of a one dimensional defect lattice within a two-dimensional confined colloidal solid}

\author{Yu-Hang Chui}
\affiliation{
Institut f\"ur Physik, Johannes-Gutenberg Universt\"at
D-55099 Mainz, Staudinger Weg 7, Germany
}
\author{Surajit Sengupta}
\affiliation{Centre for Advanced Materials, Indian Association for the Cultivation of Science,
2A \& 2B Raja S.C.  Mallik Road, Jadavpur, Kolkata, West Bengal 700 032, India
}
\affiliation{Advanced Materials Research Unit,
Satyendra Nath Bose National Centre for Basic Sciences, Block-JD, 
Sector-III, Salt Lake, Kolkata 700 098, India
}
\author{Ian K. Snook}
\affiliation{
Applied Physics, School of Applied Science, RMIT University, B. O. Box 2476V, 3001 Victoria, Australia
}
\author{Kurt Binder}
\affiliation{
Institut f\"ur Physik, Johannes-Gutenberg Universt\"at
D-55099 Mainz, Staudinger Weg 7, Germany
}
\date{\today}
\begin{abstract}
We report Monte Carlo studies of a two-dimensional soft colloidal crystal confined in a strip geometry by parallel walls. The wall-particle interaction has corrugations along the length of the strip. Compressing the crystal by decreasing the distance between the walls induces a structural transition characterized by the sudden appearance of a one-dimensional array of extended defects each of which span several lattice parameters, a ``soliton staircase''. We obtain the effective interaction between these defects. A Lindemann criterion shows that the reduction of dimensionality causes a finite periodic chain of these defects to readily melt as the temperature is raised. We discuss possible experimental realizations and speculate on potential applications.
\end{abstract} 
\pacs{74.25.Qt,61.43.Sj,83.80.Hj,05.65.+b}
\maketitle

There are many examples of condensed matter systems where extended defects in some order parameter field behave as effective ``particles'' which themselves undergo order-disorder transitions with important consequences for the properties of the original system. One can easily recall many examples such as charge\cite{11} or spin\cite{12} density waves, vortex matter\cite{13}, Skyrmions\cite{7} in fractional quantum Hall systems, domain walls in commensurate-incommensurate phases\cite{6} etc. In many of these examples, the typical size of these defects is much larger than the smallest relevant microscopic length scale. Investigation of the properties of such defect lattices requires knowledge of the effective interactions between defects which are usually difficult to measure directly in experiments. They are also difficult to obtain from computer simulations because of the large difference in length scales involved and can usually be computed only within a mean field approach and in the highly dilute limit\cite{chai}. In this Rapid Communication we describe a simple example involving extended defects in a colloidal solid\cite{8,9} where, on the other hand,  such effective interactions may be obtained to great accuracy using a relatively small system with appropriate use of finite size techniques.

\begin{figure}[ht]
\begin{center}
\includegraphics[width=3.0 in]{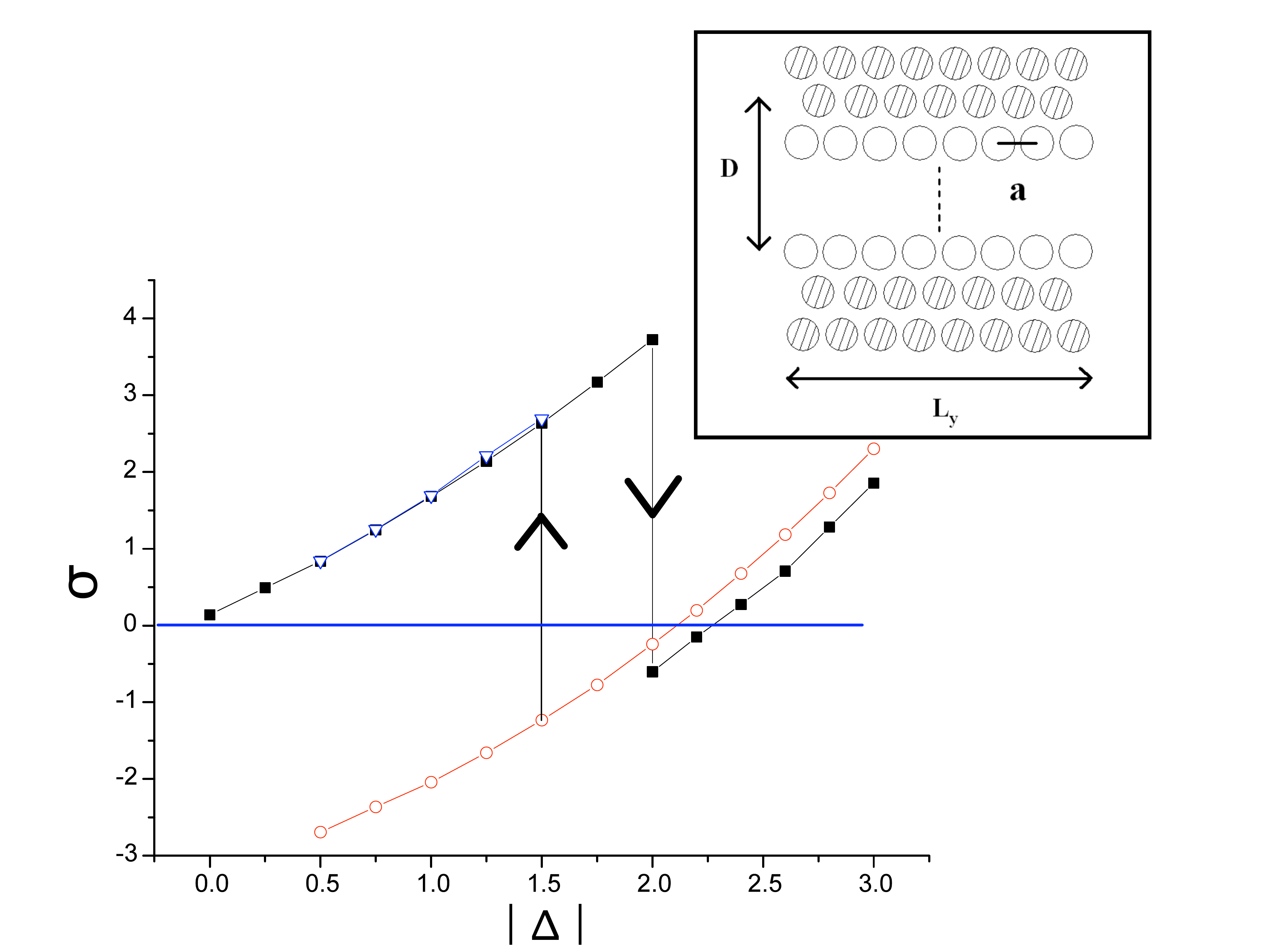}
\caption{
Internal stress $\sigma = \sigma_{xx} - \sigma_{yy}$ (in Lennard-Jones units) in the confined crystalline strip plotted vs. $\Delta$, for the case of a system started with $n_x = 30, n_y = 108$ (full symbols) and a system started with $n_x = 29, n_y = 108$ with the $108$ extra particles distributed among the $27$ inner rows, (open symbols). Curves are guides to the eye only.
Inset: A schematic sketch of our geometry: we study a system of size $D$ in $x$-direction and $L_y$ in y-direction, apply periodic boundary condition along the $y$-axis, while the boundary in the $x$-direction is created by two rows of fixed particles (shaded) on the ideal positions of a perfect triangular lattice at each side. In the fully commensurate case, $D = n_x a \sqrt{3}/2$.  
}
\label{fig1}
\end{center}
\end{figure}

In a recent work \cite{10}, we have shown that one can produce novel defect states in a colloidal crystal confined within a narrow quasi one dimensional strip by deforming it in a suitable way. We perform Monte Carlo simulations\cite{18} of a simple model solid with particles interacting with a potential\cite{16,17} $V(r) = \varepsilon (\frac{\sigma}{r})^{12}$ at a distance $r$. 
The  solid consisting of $n_y \times n_x$ unit cells with lattice parameter $a$ is confined in a channel of length $L_y = n_y a $ and width $D = (n_x - \Delta) a \sqrt{3} /2$, where $\Delta$ is a ``misfit'' parameter (Fig.\ref{fig1} (inset)). Periodic boundary conditions are assumed in the $x$ direction whereas, in the $y$ direction, the crystalline strip is confined by two fixed walls composed of two rows of immobile particles. When $\Delta = 0$, one obtains a triangular crystalline solid between the walls at zero tensile stress $\sigma$. With increasing misfit $\Delta$ (i.e. tensile strain)  $\sigma$ increases up to some critical value, where a transition occurs that reduces $n_x$ by one. At constant density, the $n_y$ extra particles of the row that disappears are added to the $n_x -3$ inner rows of the strip; the resulting average lattice spacing $a' = a (n_x -2)/(n_x-3)$ is incommensurate with the effective periodic potential due to the rows of fixed particles. This leads to the formation of a ``soliton staircase'' \cite{10} along the length of the walls, (accompanied by a pattern of standing strain waves in the crystal) according to the Frenkel-Kontorova mechanism\cite{25}. The number of solitons produced is given by $n_s = n_y/(n_x-3)$\cite{10} since each soliton contais just a single excess particle. 
Here we extend our study and investigate the structural and mechanical properties of the system, and show that the soliton superstructure in confined crystals behaves as a one dimensional system of extended ``particle''-like excitations which interact among themselves via an ``effective'' harmonic potential. We show how to extract the harmonic ``spring constant'' of this effective lattice and study the gradual melting of the soliton lattice into a Òsoliton fluidÓ caused by raising the temperature. We expect our calculations to be of direct relevance to experiments on confined colloidal crystals\cite{8,9,15}.

In order to extract properties of the defect system, we need to obtain the size and position of the individual defects. We describe below the two independent techniques that we have used for this purpose. We shall also show that the results obtained by these two methods agree with each other making us confident of our conclusions.

\begin{figure}[ht]
\begin{center}
\includegraphics[width=3.0 in]{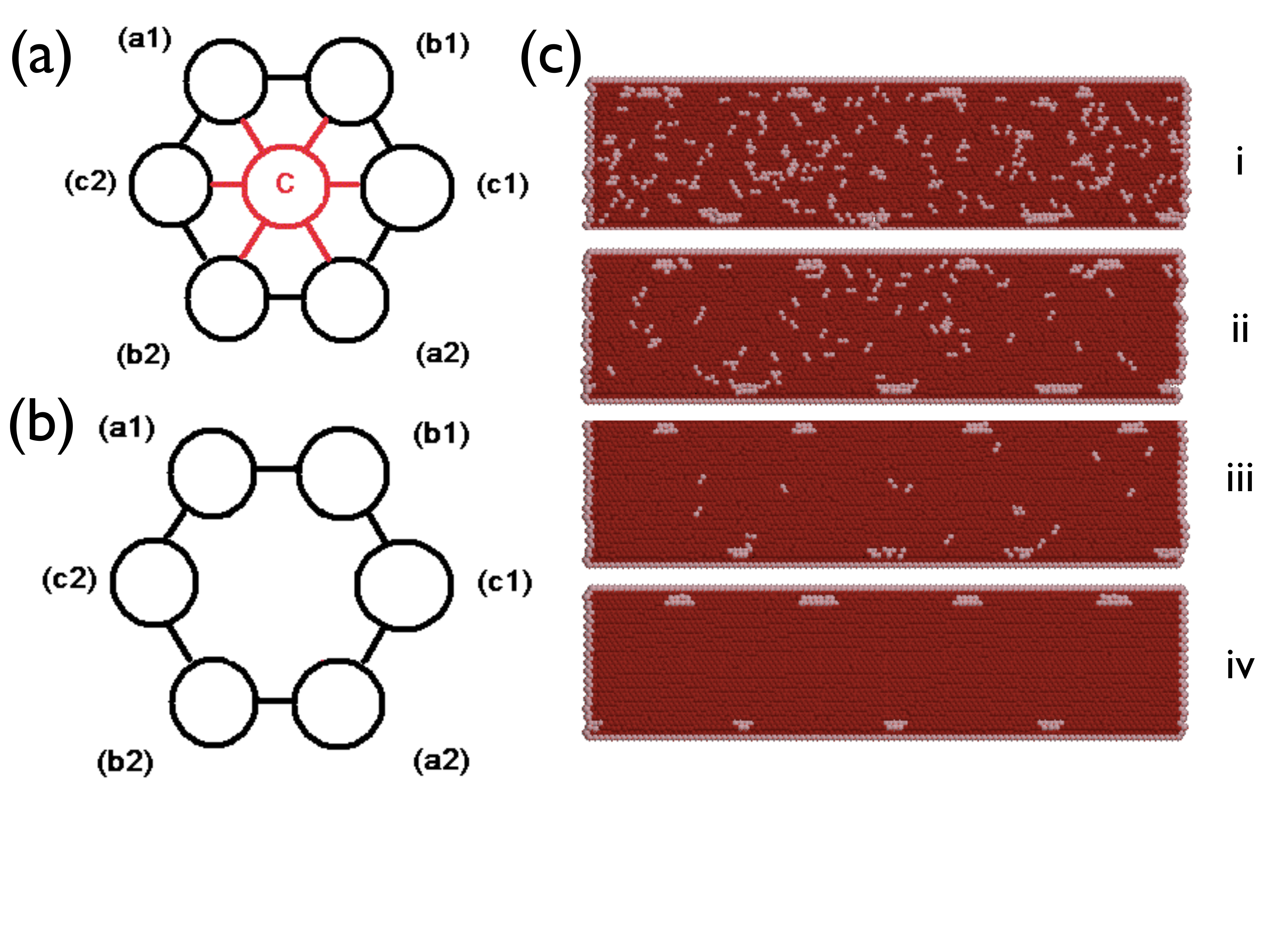}
\caption{
Characterization of the 6-membered SP rings: (a) the modified 6-membered ring where C is the central particle, (b) an SP-6-membered ring containing three antipodal pairs (a1, a2), (b1, b2), (c1, c2) the shortest paths of equal distance between (a1) and (a2) are through (b1) and (c1) or through (c2) and (b2).
(c) The results of two-dimensional ring analysis on $108 \times 30$ colloidal crystals. In each case, a defect lattice consisting of $n_s = 4 = 108/(30-3)$ solitons was stabilized at temperatures $T = 0.7 (i), 0.5 (ii), 0.3 (iii)$ and $0.1 (iv)$. Note that as the temperature is increased, random thermal fluctuations produce defects within the bulk of the strip, making the identification of the defect lattice more and more ambiguous.
}
\label{fig2}
\end{center}
\end{figure}

{\em 1. The ring method:\,\,} Here, we identify the particles which belong to a single defect using the topologically defined concept of {\em shortest path} (SP) ring structures; the procedure is discussed at length in Refs. \cite{20,21,22}, to which we refer the interested reader for details. Briefly, we classify particles as belonging either to a standard $6$-membered SP ring or to a {\em modified} $6$-membered SP ring.  A particle belongs to an SP ring if the number of bonds passed through in moving from one particle of the ring to another is the shortest among all possible paths through the network of bonds. If, in addition, every particle of the ring is also bonded to a single central particle, the particles are said to belong to a modified SP ring. By definition, particles belonging only to SP rings represent regions containing defects, while those belonging to modified SP rings represent ideal crystalline arrangement.
Once the positions of the atoms belonging to defect (soliton) locations are obtained, one can use standard cluster counting techniques to obtain the coordinates of the atoms associated with each individual defect. We observe that (1) the defects are extended structures consisting of more than one atom and (2) the number of atoms comprising a defect is more or less fixed. We can then easily obtain the center of mass coordinates of each defect (i.e. group of pink particles, see Fig. \ref{fig2}(c)).
In Fig.\ref{fig2}(c),  the red and pink colors represent the locally ideal and defective neighborhoods, respectively, inside the 2D strained colloidal crystal showing the stabilization of a soliton lattice at four different temperatures. Note that, as the temperature is increased, the location and size of the defects become ambiguous due to the presence of random thermal fluctuations.  

\begin{figure}[ht]
\begin{center}
\includegraphics[width=3.0 in]{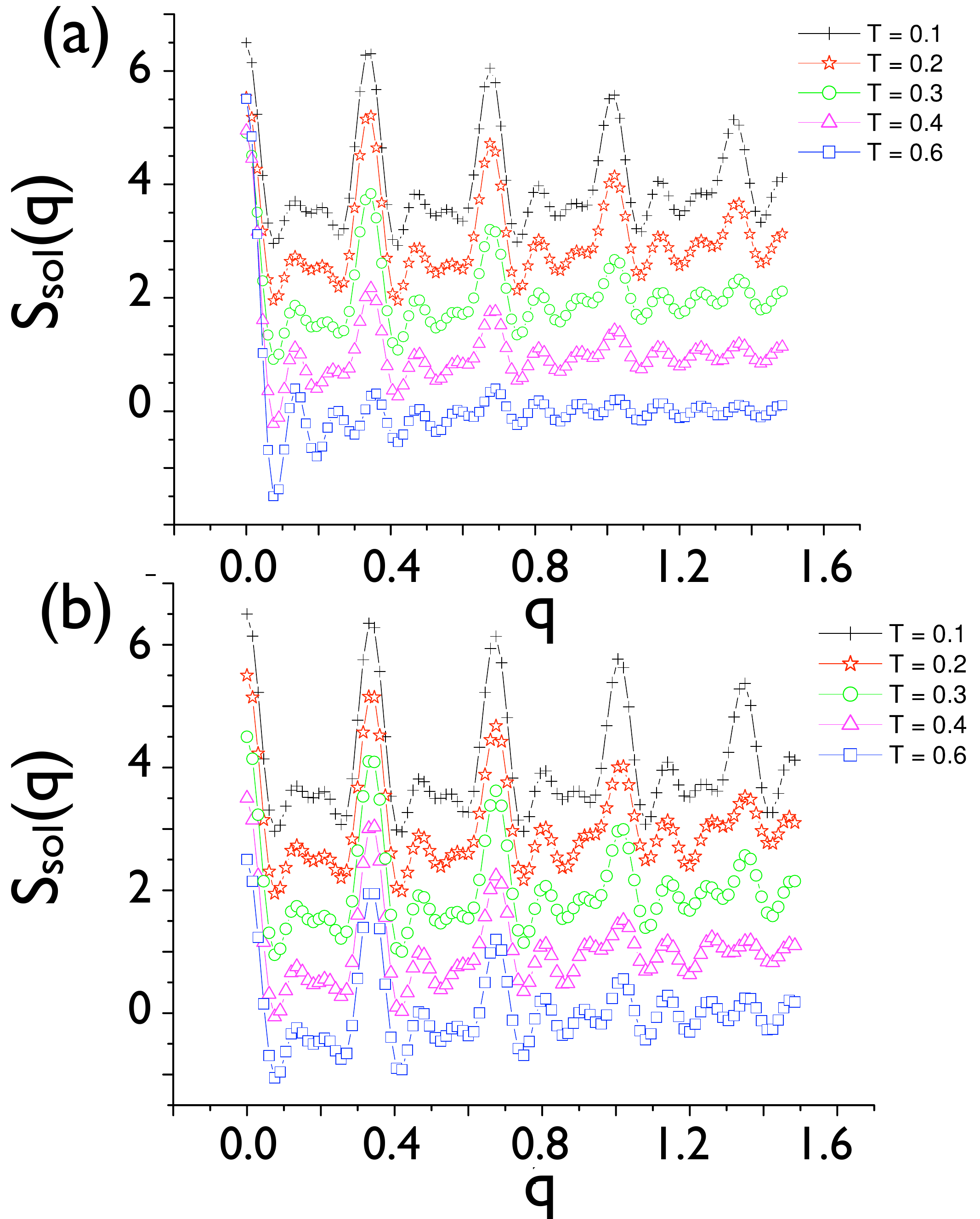}
\caption{
Structure factor of the soliton staircase $S_{sol}(q)$  vs. $ q $ at temperatures $T=0.1$($+$), 0.2($*$), 0.3 ($\circ$), 0.4 ($\Delta$) and 0.6 ($\Box$). Symbols are the actual simulation data obtained from either the ring-analysis (a) or the block analysis (b), lines are guides to the eye.  Each curve has been shifted by an unit distance in the vertical direction for clarity. All data are for the system with $L = n_y a$  with $n_y = 108$ , and  $n_s =6$ solitons present after a transition from $n_x = 20$ to $n_x=19$. Note that both (a) and (b) give comparable results.
}
\label{fig3}
\end{center}
\end{figure} 

{\em 2. The block method:\,}  It is important to make sure that the soliton positions that we extract from the simulation, are not affected seriously by the method used to analyze these configurations. Therefore we have used another method, viz. the Òblocking methodÓ, to obtain an independent estimation. Accordingly, we introduce a block length $L_b = n_b a$, where $a$ is the lattice constant of the undeformed lattice. We choose $n_b= 8-12$ and hence $L_b$ such that $n_b \gg 1$ but $L_b$ still clearly less than the expected value of  $\Delta y_0$ . By moving this coarse-graining block along a row in the $y$-direction (in steps of $a$), we can count how many particles $n$ actually fall inside a block. If we work at low enough temperatures, where the mean-square displacement of the particles on the lengths $L_b$ is still clearly much less than $a^2$, we obtain $n = n_b$ if no soliton core falls into the block, while we obtain $n = n_b + 1$ if a soliton core falls into the block. Calculating then the center of mass of a cluster of adjacent blocks with $n = n_b + 1$ then yields an alternative estimate for the position of a soliton in a system configuration.

It is now straightforward to obtain both the structure factor $S_{sol}(q) = \frac {1}{n_x} \sum_{{l l'}} \langle \exp [ i\,q\,(y_{l} - y_{l'})] \rangle $ and the probability distribution $P(\Delta y_0)$ of $\Delta y_0$ the distance between the centre of mass positions $y_l, y_{l+1}$ of neighboring solitons $l, l+1$. The sum runs over the positions of all solitons in a row, and the result is averaged over all rows over both boundaries of the system apart from the statistical average over many independent particle configurations). While at $T = 0$ in the limit $L_y $ and $n_s \to \infty$ the soliton lattice would cause sharp Bragg peaks, for all $T > 0$ the soliton system is expected to have a liquid-like structure factor. In fact, assuming that the interaction between neighboring solitons is harmonic, 
\begin{equation}
S_{sol}(q) = \frac {1}{n_x} \sum_{{l l'}} \exp[iq(l - l') \Delta y_0] \exp (-\frac{1}{2} q^2 \langle (U_l - U_{l'})^2 \rangle )
\label{sq1}
\end{equation}

Each soliton is described as an effective point particle of mass M, position $y_l$ and conjugate momentum $\Pi_l$. The Hamiltonian for the harmonic chain is given by, 
\begin{equation}
H_{sol} = \frac{1}{2} \sum_l [\Pi^2_l/M + M C^2 (y_{l+1} - y_l - \Delta y_0)^2/(\Delta y_0)^2]
\label{hamil}
\end{equation}
Where the parameter $C$ plays the role of a sound velocity. From Eq. \ref{hamil} one obtains the correlation function of the mean square displacements  $U_l = y_l - l\,\Delta y_0 $ as\cite{27}
\begin{eqnarray}
\langle(U_l - U_0)^2\rangle & = & l (\Delta y_0)^2 k_B T/(MC^2) \nonumber \\ 
                                            & = & l \delta^2 
\label{disp}
\end{eqnarray}
Here $\delta$ characterizes the local displacement, $\delta^2 = \langle (U_{l+1}-U_l)^2\rangle = \langle(\Delta y_l )^2 \rangle$. The obvious interpretation of Eq. \ref{disp} is that the relative displacements $\Delta y_l$  at each index $l$  of the one-dimensional soliton lattice add up in a random-walk-like fashion\cite{17}. From Eqs. \ref{sq1},\ref{disp}, one may derive
$S_{sol}(q)$  in one dimension, for $n_x \to \infty$\cite{27,28} exactly. However, our data corresponds to very small $n_s \sim 4-6$. Nevertheless, following Ref. \cite{28} one can evaluate Eq. \ref{sq1} for finite $n_s$, using $\delta$  and $\Delta y_0$ as parameters,
\begin{equation}
S_{sol}(q) = \frac {1}{2} \sum_{l=1}^{n_s-1} \cos(ql\Delta y_0) \exp (-\frac{1}{2} q^2 \delta^2 l )
\label{sq4}
\end{equation}
 			
As long as  $\delta = \langle (\Delta y)^2 \rangle^{1/2} \ll \Delta y_0$, the one-dimensional correlation extends over many solitons, and the term Òsoliton latticeÓ still is in a sense meaningful; when $\delta$ is no longer much smaller than $\Delta y_0$, however, the system rather should be described as a Òsoliton liquidÓ. As is well known, the melting of a one-dimensional crystal is a continuous transition ($\xi \sim 1/T$). If one nevertheless defines\cite{22} an effective melting temperature $T_m^{sol}$ for one-dimensional systems by arbitrarily requiring that their Lindemann parameter $\delta^2/\Delta y_0 < 0.01$,  one would obtain $k_B T_m^{sol} = 0.01 MC^2$ as the temperature scale that controls the ÒmeltingÓ of the soliton lattice. Though clearly the melting of the soliton lattice is far from a sharp thermodynamic phase transition, this already suggests that the soliton lattice may melt at rather low temperature, far below the melting temperature $T_m Å 1.35$ of the bulk two-dimensional crystal at the chosen density\cite{16}.
	 		
Fig. \ref{fig3} shows simulation data for $S_{sol}(q)$ vs. $q$ at various temperatures. We note that at $T = 0.1$ indeed the peaks of $S_{sol}(q)$ are already rather sharp, while for $T > 0.3$ the structure factor clearly has the character of a fluid. The nature of the curves are well represented by the form given in Eq.\ref{sq4} and we may obtain a value for the (only) parameter $\delta$ by fitting the data. However, we use the more accurate procedure discussed below. 

\begin{figure}[ht]
\begin{center}
\includegraphics[width=4.0 in]{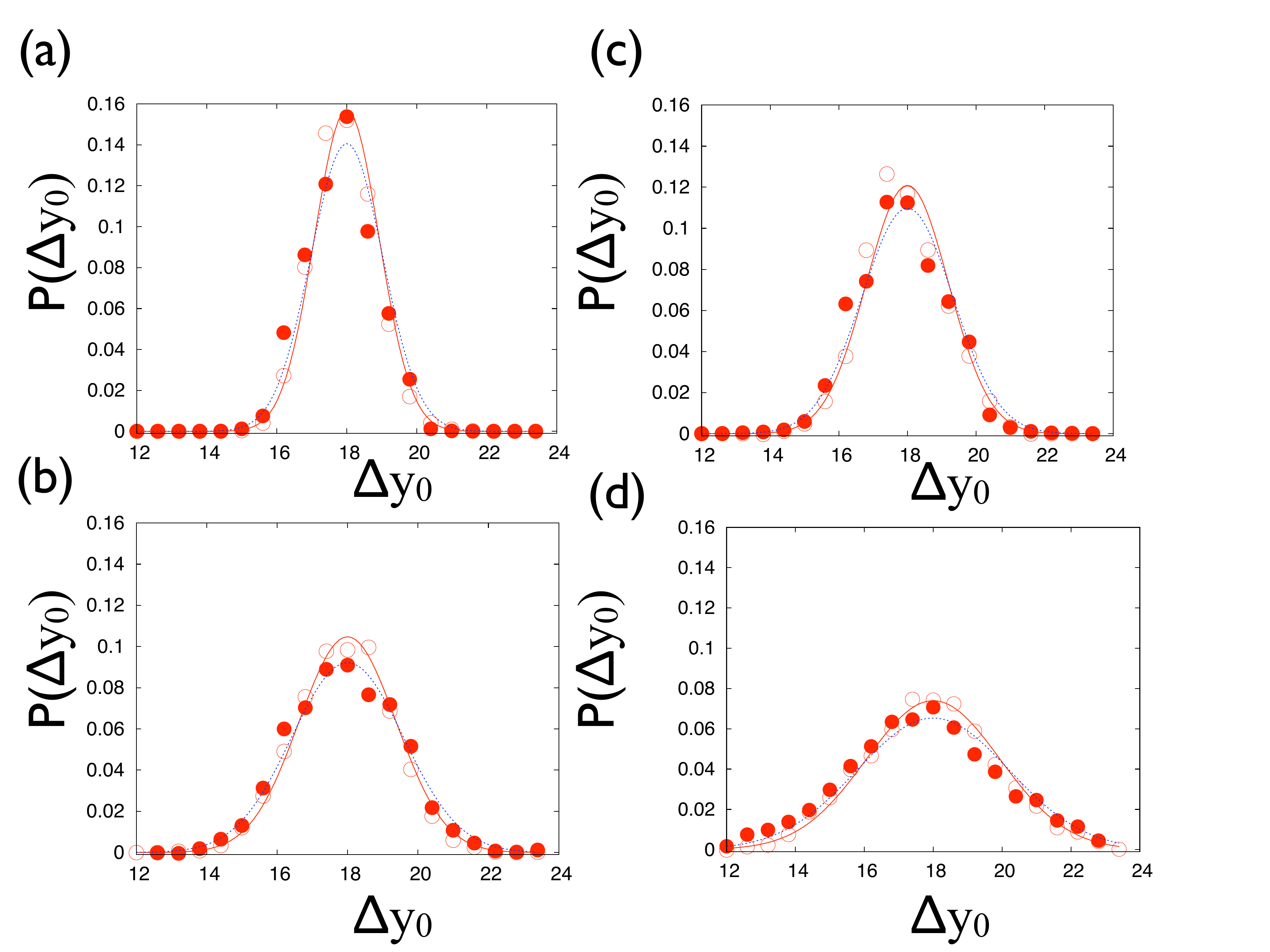}
\caption{
Probability distribution $P(\Delta y_0)$  versus $\Delta y_0$, for $T=0.1$(a), $0.2$(b), $0.3$ (c) and $0.6$ (d). Open and closed symbols are the results from the block and the 2d ring analysis method respectively. The lines show the corresponding gaussian fits.
}
\label{fig4}
\end{center}
\end{figure}

\begin{figure}[ht]
\begin{center}
\includegraphics[width=1.5 in]{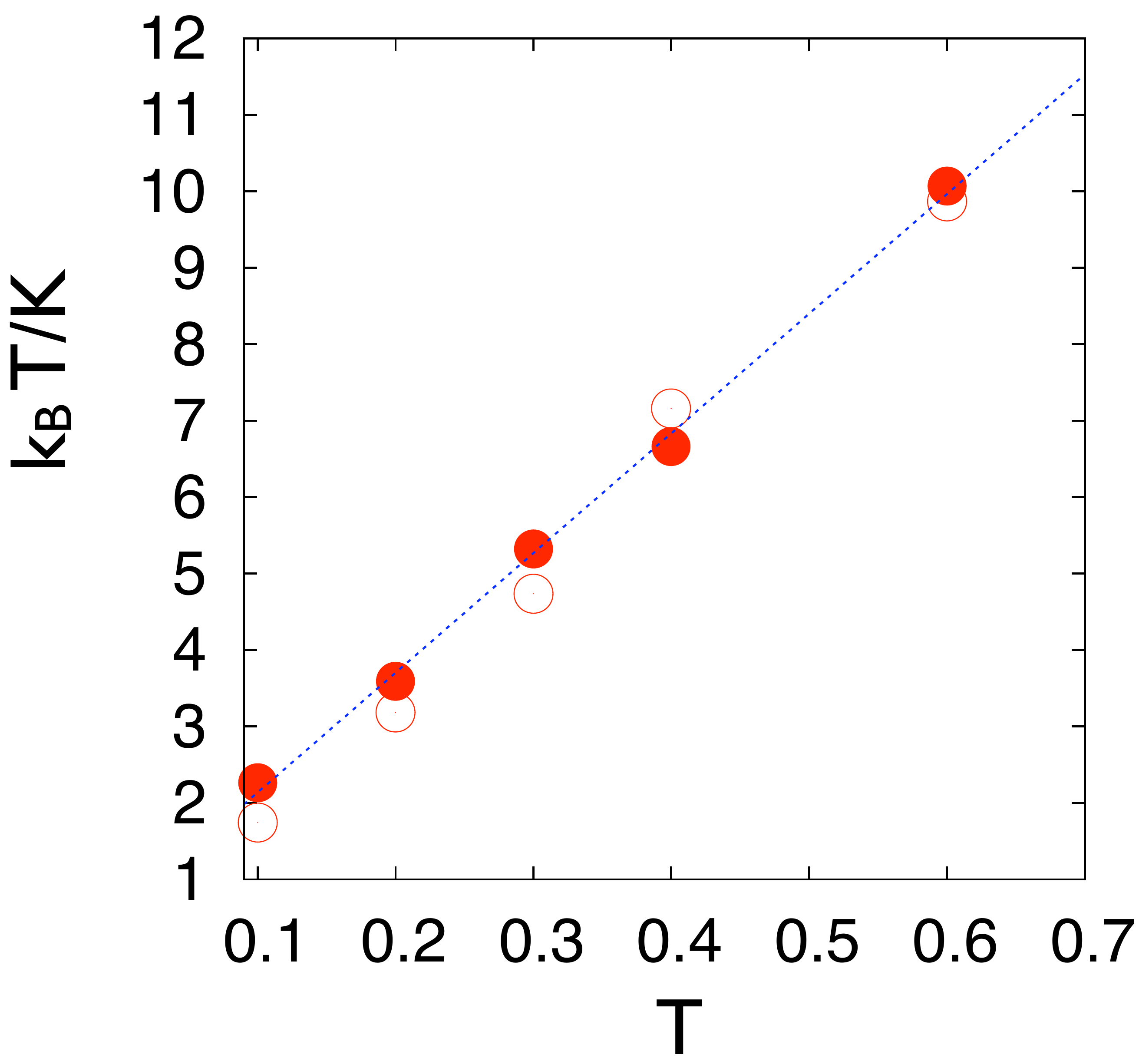}
\caption{
Plot of  $k_B T /K$, as extracted from $P(\Delta y_0)$, versus $T$. Symbols have the same meaning as in Fig. 4. The straight line is a fit to all data, yielding $K=7.3 \pm 0.3$.
}
\label{fig5}
\end{center}
\end{figure}

We now compare the probability distribution $P(\Delta y_0)$ of the distance between neighboring solitons in the lattice from both the ring analysis method and this blocking method. Fig. \ref{fig4} shows that both methods of analyzing the configurations to identify where the solitons are, agree almost perfectly with each other, and moreover $P(\Delta y_0)$ is nicely described by a Gaussian, $P(\Delta y_0)= A \exp(-\frac{1}{2} \,b\,(\Delta y - \Delta y_0)^2)$  where $A$ is a constant ensuring normalization, $\Delta y_0 = 18 = n_y/n_s = 108/6$ has been used to be consistent with our choice of $L$ and $D$ (an unconstrained fit gives $\Delta y_0 = ,17.866 \pm 0.023$ !) so the only fit parameter is $b = \langle \Delta y^2 \rangle^{-1}$. From Eq. \ref{hamil} it is obvious that the quantity $K = MC^2/(\Delta y_0)^2$ plays the role of 
an elastic constant such that $b = K/k_B T$.  Fig.\ref{fig5} therefore plots $k_B T/K $ vs. $T$, to test to what extent $K$ is independent of temperature. Finally, we note that the Lindemann ratio $0.01$ is reached at an effective melting temperature $T_m^{sol} = 0.44$. This value is compatible with the direct observation of the effective melting temperature obtained from the structure factor.

In our previous studies\cite{10}, we briefly reported our discovery of the soliton staircases and the strain wave patterns in confined soft two-dimensional colloidal crystals of different sizes. In the present paper, we obtain accurate effective interactions between the solitions and report a gradual melting of the soliton superlattice in analogy to that of harmonic chains. We believe that our studies would be useful in designing experimental colloidal systems where such defect lattices in narrow channels may be stabilized. These structures, which have some resemblance to vortex matter in channels,\cite{30} may have interesting optical and transport properties. Work along these lines are in progress. 

\section{acknowledgement}
This research was partially supported by the Deutsche Forschungsgemeinschaft Project TR6/C4. Yu-Hang Chui would like to thank RMIT University, Australia for the hospitality during his academic visits. SS thanks DST, Govt. of India for support.


\begin{thebibliography}{10}
\bibitem{11} R. E. Thorne, Physics Today {\bf 49}, 42 (1996).
\bibitem{12} S. Brown and G. Gruner, Scientific American {\bf 270}, 50 (1994).
\bibitem{13} G. Blatter, M. V. Feigel'man, V. B. Geshkenbein, A. I. Larkin, and V. M. Vinokur, Rev. Mod. Phys. {\bf 66}, 1125 (1994) 
\bibitem{7}S.L.Sondhi, A.Karlshede, S.A.Kivelson and E.H.Rezayi, Phys.Rev.B {\bf 47}, 16419 (1993).
\bibitem{6} 
F.F Abraham, W.E. Rudge, D.J. Auerbach and S.W. Koch, Phys. Rev. Letters, 52, 445 (1984)
J. Villain in ÒOrdering in Strongly Fluctuating Condensed Matter SystemsÓ, Ed. T. Riste, Plenum, New York, (1980)
M. Mardar and A.N. Berker, Phys. Rev. Letters 48 1552 (1982)
\bibitem{chai}P. M. Chaikin and T. C. Lubensky, {\em Principles of condensed matter physics}, 
 (Cambridge University Press, Cambridge, England, 1995).
\bibitem{8} A. Blaaderen, Progr. Colloid Polym. Sci. 104, 59 (1997).
\bibitem{9} K. Zahn and G. Maret, Phys. Rev. Lett. 85, 3656 (2000); W. Poon, Science 304, 830 (2004).
\bibitem{10} Y.-H. Chui, S. Sengupta and K. Binder, Europhys. Lett. 83, 58004 (2008).
\bibitem{18}  K. Binder, Rep. Progr. Phys. 60, 487 (1997).
\bibitem{16} K. Bagchi, H. C. Andersen and W. Swope, Phys. Rev. E 53, 3794 (1996).
\bibitem{17} A. Ricci, P. Nielaba, S. Sengupta and K. Binder, Phys. Rev. E 75, 011405 (2007).
\bibitem{25} O. M. Braun and Y. S. Kivshar, The Frenkel-Kontorova-Model: Concepts, Methods and Applications (Springer, Berlin, 2004)
\bibitem{15} M. Koppl., P. Henseler, A. Erbe, P. Nielaba, and P. Leiderer, Phys. Rev. Lett. 97, 208302 (2006). 
\bibitem{20} B. OÕMalley, Ph.D thesis, RMIT University (2001); B. OÕMalley and I. K. Snook, Phys. Rev. Lett. 90, 085702 (2003); B. OÕMalley and I. Snook, J. Chem. 
Phys. 123, 054511 (2005).
\bibitem{21} R. J. Rees, Ph. D thesis, RMIT University (2004). 
\bibitem{22} D. S. Franzblau, Phys. Rev. B 44, 4925 (1991).

\bibitem{27} V. J. Emery and J. D. Axe, Phys. Rev. Lett. 40, 1507 (1978)
\bibitem{28} G- Radons, J. Keller and T. Geisel, Z. Phys.B-Condens. Matter 50, 289 (1983)
\bibitem{30} See N. Kokubo, R. Besseling, and P. H. Kes, Phys. Rev. B69, 064504 (2004) and references therein.
\end{thebibliography}
\end{document}